\documentclass[aps,prl,preprint,tightenlines,superscriptaddress,showpacs,byrevtex]{revtex4}
\usepackage{dcolumn}  

\usepackage{graphicx} 
\usepackage{color} %
\usepackage[hypertex]{hyperref} 
\RequirePackage{xspace}
\usepackage{relsize}
\def\babar{\mbox{\slshape B\kern-0.1em{\smaller A}\kern-0.1em B\kern-0.1em{\smaller A\kern-0.2em R}}}
\def\qbar  {\ensuremath{\overline q}\xspace}
\def\KS    {\ensuremath{K^0_{\scriptscriptstyle S}}\xspace} 
\def\KL    {\ensuremath{K^0_{\scriptscriptstyle L}}\xspace} 
\def\Dbar  {\kern 0.2em\overline{\kern -0.2em D}{}\xspace}
\def\Dzb   {\ensuremath{\Dbar^0}\xspace}
\def\Bbar  {\kern 0.2em\overline{\kern -0.2em B}{}\xspace}
\mathchardef\Upsilon="7107
\def\Y#1S{\ensuremath{\Upsilon{(#1S)}}\xspace}
\def\cm    {\ensuremath{{\rm \,cm}}\xspace}
\def\invfb {\ensuremath{\mbox{\,fb}^{-1}}\xspace}
\def\CP    {\ensuremath{C\!P}\xspace}
\def\nb    {\ensuremath{C_{N\!B}}\xspace}
\def\nblow{\ensuremath{C_{N\!B,{\rm low}}}\xspace}
\def\nbhig{\ensuremath{C_{N\!B,{\rm high}}}\xspace}
\def\nbprim{\ensuremath{C^{\prime}_{N\!B}}\xspace}
\def\nbprimi{\ensuremath{C^{\prime\,i}_{N\!B}}\xspace}

\newcommand{\stat}{\ensuremath{\mathrm{(stat)}}\xspace}
\newcommand{\syst}{\ensuremath{\mathrm{(syst)}}\xspace}
\newcommand{\gev}{\ensuremath{\mathrm{\,Ge\kern -0.1em V}}\xspace}
\newcommand{\mev}{\ensuremath{\mathrm{\,Me\kern -0.1em V}}\xspace}
\newcommand{\gevc}{\ensuremath{{\mathrm{\,Ge\kern -0.1em V\!/}c}}\xspace}
\newcommand{\mevc}{\ensuremath{{\mathrm{\,Me\kern -0.1em V\!/}c}}\xspace}
\newcommand{\gevcc}{\ensuremath{{\mathrm{\,Ge\kern -0.1em V\!/}c^2}}\xspace}
\newcommand{\mevcc}{\ensuremath{{\mathrm{\,Me\kern -0.1em V\!/}c^2}}\xspace}
\newcommand{\etapr}{\ensuremath{\eta^{\prime}}\xspace}

\begin{document}

\resizebox{!}{3cm}{\includegraphics{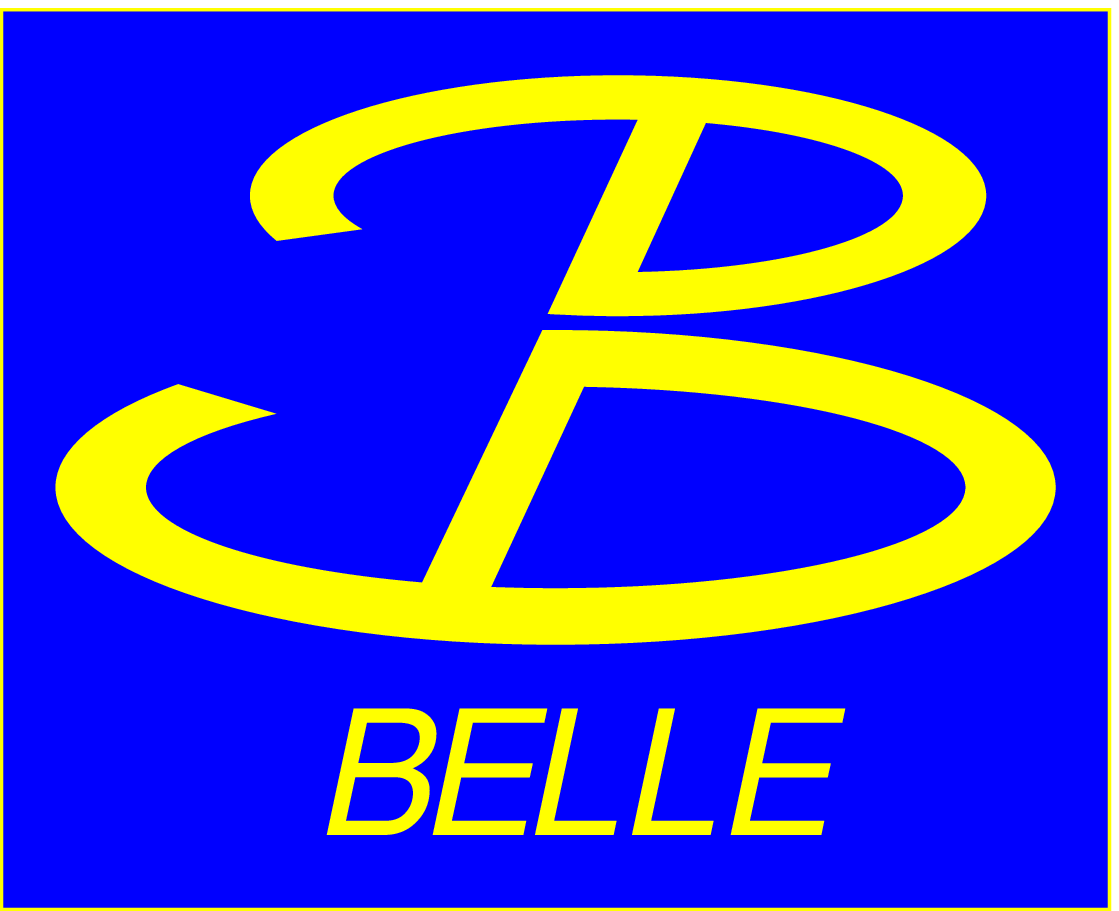}}

\vspace*{-2.5\baselineskip}

\preprint{\vbox{ \hbox{}
\hbox{}
\hbox{}
\hbox{BELLE Preprint 2014-13}
\hbox{KEK Preprint 2014-22}
}}

\title{\quad\\[1.0cm] Observation of the decay {\boldmath $B^0\to\etapr K^{*}(892)^{0}$}}


\noaffiliation
\affiliation{University of the Basque Country UPV/EHU, 48080 Bilbao}
\affiliation{Beihang University, Beijing 100191}
\affiliation{Budker Institute of Nuclear Physics SB RAS and Novosibirsk State University, Novosibirsk 630090}
\affiliation{Faculty of Mathematics and Physics, Charles University, 121 16 Prague}
\affiliation{Chonnam National University, Kwangju 660-701}
\affiliation{University of Cincinnati, Cincinnati, Ohio 45221}
\affiliation{Deutsches Elektronen--Synchrotron, 22607 Hamburg}
\affiliation{Justus-Liebig-Universit\"at Gie\ss{}en, 35392 Gie\ss{}en}
\affiliation{Gifu University, Gifu 501-1193}
\affiliation{The Graduate University for Advanced Studies, Hayama 240-0193}
\affiliation{Hanyang University, Seoul 133-791}
\affiliation{University of Hawaii, Honolulu, Hawaii 96822}
\affiliation{High Energy Accelerator Research Organization (KEK), Tsukuba 305-0801}
\affiliation{IKERBASQUE, Basque Foundation for Science, 48011 Bilbao}
\affiliation{Indian Institute of Technology Bhubaneswar, Satya Nagar 751007}
\affiliation{Indian Institute of Technology Guwahati, Assam 781039}
\affiliation{Indian Institute of Technology Madras, Chennai 600036}
\affiliation{Institute of High Energy Physics, Chinese Academy of Sciences, Beijing 100049}
\affiliation{Institute of High Energy Physics, Vienna 1050}
\affiliation{Institute for High Energy Physics, Protvino 142281}
\affiliation{INFN - Sezione di Torino, 10125 Torino}
\affiliation{Institute for Theoretical and Experimental Physics, Moscow 117218}
\affiliation{J. Stefan Institute, 1000 Ljubljana}
\affiliation{Kanagawa University, Yokohama 221-8686}
\affiliation{Institut f\"ur Experimentelle Kernphysik, Karlsruher Institut f\"ur Technologie, 76131 Karlsruhe}
\affiliation{Department of Physics, Faculty of Science, King Abdulaziz University, Jeddah 21589}
\affiliation{Korea Institute of Science and Technology Information, Daejeon 305-806}
\affiliation{Korea University, Seoul 136-713}
\affiliation{Kyungpook National University, Daegu 702-701}
\affiliation{\'Ecole Polytechnique F\'ed\'erale de Lausanne (EPFL), Lausanne 1015}
\affiliation{Faculty of Mathematics and Physics, University of Ljubljana, 1000 Ljubljana}
\affiliation{Luther College, Decorah, Iowa 52101}
\affiliation{University of Maribor, 2000 Maribor}
\affiliation{Max-Planck-Institut f\"ur Physik, 80805 M\"unchen}
\affiliation{School of Physics, University of Melbourne, Victoria 3010}
\affiliation{Moscow Physical Engineering Institute, Moscow 115409}
\affiliation{Moscow Institute of Physics and Technology, Moscow Region 141700}
\affiliation{Graduate School of Science, Nagoya University, Nagoya 464-8602}
\affiliation{Kobayashi-Maskawa Institute, Nagoya University, Nagoya 464-8602}
\affiliation{Nara Women's University, Nara 630-8506}
\affiliation{National Central University, Chung-li 32054}
\affiliation{National United University, Miao Li 36003}
\affiliation{Department of Physics, National Taiwan University, Taipei 10617}
\affiliation{H. Niewodniczanski Institute of Nuclear Physics, Krakow 31-342}
\affiliation{Niigata University, Niigata 950-2181}
\affiliation{Osaka City University, Osaka 558-8585}
\affiliation{Pacific Northwest National Laboratory, Richland, Washington 99352}
\affiliation{Peking University, Beijing 100871}
\affiliation{University of Pittsburgh, Pittsburgh, Pennsylvania 15260}
\affiliation{University of Science and Technology of China, Hefei 230026}
\affiliation{Seoul National University, Seoul 151-742}
\affiliation{Soongsil University, Seoul 156-743}
\affiliation{Sungkyunkwan University, Suwon 440-746}
\affiliation{School of Physics, University of Sydney, NSW 2006}
\affiliation{Department of Physics, Faculty of Science, University of Tabuk, Tabuk 71451}
\affiliation{Tata Institute of Fundamental Research, Mumbai 400005}
\affiliation{Excellence Cluster Universe, Technische Universit\"at M\"unchen, 85748 Garching}
\affiliation{Toho University, Funabashi 274-8510}
\affiliation{Tohoku University, Sendai 980-8578}
\affiliation{Department of Physics, University of Tokyo, Tokyo 113-0033}
\affiliation{Tokyo Institute of Technology, Tokyo 152-8550}
\affiliation{Tokyo Metropolitan University, Tokyo 192-0397}
\affiliation{University of Torino, 10124 Torino}
\affiliation{CNP, Virginia Polytechnic Institute and State University, Blacksburg, Virginia 24061}
\affiliation{Wayne State University, Detroit, Michigan 48202}
\affiliation{Yamagata University, Yamagata 990-8560}
\affiliation{Yonsei University, Seoul 120-749}
  \author{S.~Sato}\affiliation{Niigata University, Niigata 950-2181} 
  \author{Y.~Yusa}\affiliation{Niigata University, Niigata 950-2181} 
  \author{G.~B.~Mohanty}\affiliation{Tata Institute of Fundamental Research, Mumbai 400005} 
  \author{A.~Abdesselam}\affiliation{Department of Physics, Faculty of Science, University of Tabuk, Tabuk 71451} 
  \author{I.~Adachi}\affiliation{High Energy Accelerator Research Organization (KEK), Tsukuba 305-0801}\affiliation{The Graduate University for Advanced Studies, Hayama 240-0193} 
  \author{H.~Aihara}\affiliation{Department of Physics, University of Tokyo, Tokyo 113-0033} 
  \author{S.~Al~Said}\affiliation{Department of Physics, Faculty of Science, University of Tabuk, Tabuk 71451}\affiliation{Department of Physics, Faculty of Science, King Abdulaziz University, Jeddah 21589} 
  \author{D.~M.~Asner}\affiliation{Pacific Northwest National Laboratory, Richland, Washington 99352} 
  \author{T.~Aushev}\affiliation{Institute for Theoretical and Experimental Physics, Moscow 117218} 
  \author{R.~Ayad}\affiliation{Department of Physics, Faculty of Science, University of Tabuk, Tabuk 71451} 
% ...
\collaboration{The Belle Collaboration}


\begin{abstract}
We report a search for charmless hadronic decays of neutral $B$ mesons to $\etapr K^{*}(892)^{0}$. The
results are based on a $711\invfb$ data sample that contains $772\times 10^6 B\Bbar$ pairs, 
collected at the $\Y4S$ resonance with the Belle detector at the KEKB asymmetric energy $e^+e^-$
collider. We observe the decay for the first time with a significance of $5.0$ standard deviations
and obtain its branching fraction ${\cal B}[B^0\to\etapr K^{*}(892)^{0}]=[2.6\pm0.7\stat\pm0.2
\syst]\times 10^{-6}$. We also measure the $\CP$ violating asymmetry as ${\cal A}_{\CP}[B^0\to\etapr 
K^{*}(892)^{0}]=-0.22\pm0.29\stat\pm0.07\syst$.
\end{abstract}


\maketitle
\tighten

{\renewcommand{\thefootnote}{\fnsymbol{footnote}}}
\setcounter{footnote}{0}

Two-body charmless decays of $B$ mesons are known to be a powerful probe for testing the standard
model (SM) predictions as well as to search for new physics~\cite{charmless}. Decays to final
states containing $\eta$ and $\etapr$ mesons exhibit a distinct pattern of interferences among
the dominant contributing amplitudes and are also sensitive to a potentially large flavor-singlet
contribution~\cite{flavor}. 

Owing to the $\eta$-$\eta'$ mixing, $b \to s$ penguin and $b \to u$ tree processes 
contribute to charmless $B$ decays with an $\eta$ or $\eta'$ in the final state \cite{brokenSU3}.
The interference of those processes is constructive for the $\eta' K$ and 
$\eta K^*$ final states, whereas it is destructive for $\eta K$ and $\eta' K^*$. Therefore, the 
$B \to \eta K$ and $B \to \eta' K^*$ decays are suppressed and thus provide a good test bed to search for 
possible contributions from new physics that could be manifested in the loop diagram. 
The destructive penguin amplitude could also interfere with the small $b \to u$ tree diagram, 
giving rise to a large direct $CP$ violation. 
Recent measurements in $B \to \eta K$ from $\babar$ \cite{BABAR-CPVetaK} and
Belle \cite{Belle-CPVetaK} seem to confirm this picture. 
Direct $CP$ violation in the $B \to \eta' K^*$ decay has not yet been probed, which  
constitutes a good sample to test the aforementioned interference scheme 
to expose new physic contributions for the $\eta^({}'{}^) K^{(*)}$ system. 
Furthermore, the study of time-dependent $CP$ asymmetry in $B^0 \to \eta' K^{*}(892)^{0}$, $K^{*}(892)^{0} \to K^0 \pi^0$ 
can add useful information to an existing intriguing effect seen in the loop-dominated $b \to s q \bar{q}~(q = u,d,s)$ 
decays compared to the tree-level $b \to c \bar{c} s$ transition \cite{b2s-timedep1, b2s-timedep2, b2s-timedep3}.

The decay $B^0\to\etapr K^{*}(892)^{0}$ has been studied extensively
within the framework of perturbative QCD~\cite{QCD}, QCD factorization~\cite{QCDF}, soft collinear
effective theory~\cite{SCET} as well as $SU(3)$ flavor symmetry~\cite{SU3}, and predicted branching fractions
are in the range (1.2$-$6.3)$\times 10^{-6}$. In the past, both Belle~\cite{belle}
and $\babar$~\cite{babar} have searched for $B^0\to\etapr K^{*}(892)^{0}$ with 
the latter reporting the first evidence with a significance of $4.0$ standard
deviations ($\sigma$).

The results reported herein are based on a data sample containing $772\times 10^{6}$ $B\Bbar$ pairs collected at
the $\Y4S$ resonance with the Belle detector~\cite{Belle_detector} at the KEKB asymmetric energy
 $e^+e^-$ ($3.5$ on $8.0\gev$) collider~\cite{KEKB}. The Belle detector consists of six nested 
sub-detectors: a silicon vertex detector (SVD), a 50-layer central drift chamber (CDC), an array
of aerogel threshold Cherenkov counters (ACC), a barrel-like arrangement of time-of-flight
scintillation counters (TOF), a CsI(Tl) crystal-based electromagnetic calorimeter (ECL), and a
multilayer structure of resistive plate counters and iron plates to detect $\KL$ mesons and
muons (KLM). All but the KLM are located inside a $1.5$\,T
solenoidal magnetic field. Two inner-detector configurations were used: a $2.0\cm$ beampipe and
a three-layer SVD for the first sample of $152\times 10^6 B\Bbar$ pairs, while a $1.5\cm$
beampipe, a four-layer SVD and a small-cell CDC for the remaining $620\times 10^6
B\Bbar$ events~\cite{svd2}. The latter sample has been reprocessed with an improved track
reconstruction algorithm, which significantly increased the signal reconstruction efficiency.

We reconstruct $B^0\to\etapr K^{*}(892)^{0}$ candidates from the subsequent decay channels $\etapr\to
\eta\pi^+\pi^-$, $\eta\to\gamma\gamma$ and $K^{*}(892)^{0}\to K^+\pi^-$. 
Since the background contribution in $\etapr\to\rho\gamma$ is significantly larger than 
in $\etapr\to\eta\pi^+\pi^-$, the former decay channel is not considered in our study. 
Because of a low expected signal yield and a poor signal-to-noise ratio, we do not reconstruct $K^*(892)^0 \to K^0\pi^0$.
Consequently, time-dependent $CP$ violation in $B^0\to\etapr K^{*}(892)^{0}$ is not treated in this paper.

Charged track candidates are required to
have a transverse momentum greater than $0.1\gevc$ and an impact parameter with respect to the
interaction point (IP) of less than $0.2\cm$ in the $r$--$\phi$ plane and $5.0\cm$ along the $z$ axis.
Here, the $z$ axis is defined as the direction opposite the $e^+$ beam. To distinguish charged
kaons from pions, we use a likelihood ratio $R_{K/\pi} = {\cal L}_K / ({\cal L}_K+ {\cal L}_\pi)$,
where ${\cal L}_K$ (${\cal L}_\pi$) denotes the likelihood for a track being a kaon (pion) and is
calculated using specific ionization in the CDC, time-of-flight information from the TOF and the
number of photoelectrons from the ACC. 
Based on this quantity, we select charged tracks to reconstruct the $\etapr$ and $K^{*}(892)^{0}$ candidates.
Since few fake $\etapr$ arising from misidentification of pions are expected, 
we apply looser conditions for pion candidates in the $\etapr$ reconstruction. 
Typical average efficiencies and fake rates in the entire momentum range 
for the kaon and pion selections are 90\% and 5\%, respectively. When applying the looser selection 
for pions, these are 95\% and 10\%, respectively.
To reconstruct $\eta$ candidates, photons originating from their decays are
required to have an energy greater than $0.1\gev$ in the ECL and an energy balance---the ratio
between the absolute difference and the sum of the two photon energies---of less than $0.9$. The $\eta$
candidates must satisfy $0.510\gevcc< M_\eta<0.575\gevcc$, corresponding to $\pm2.5\sigma$ 
around the nominal $\eta$ mass~\cite{PDG}. The $\etapr$ candidates are 
required to satisfy $0.950\gevcc <M_{\etapr}<0.965\gevcc$, corresponding to $\pm2.5\sigma$ around 
the nominal $\etapr$ mass~\cite{PDG}. Finally, the $K^{*}(892)^{0}$ candidates must have  
$0.820\gevcc<M_{K^{*}(892)^{0}} <0.965\gevcc$.

We identify $B$ candidates using two kinematic variables: the beam-energy constrained mass, $M_{\rm bc}
\equiv\sqrt{E^{2}_{\rm beam}-|\sum_{i}\vec{p}_{i}|^{2}}$, and the energy difference, $\Delta E\equiv\sum_{i}E_{i}-
E_{\rm beam}$, where $E_{\rm beam}$ is the beam energy, and $\vec{p}_{i}$ and $E_{i}$ are the momentum and
energy, respectively, of the $i$-th daughter of the reconstructed $B$ candidate in the $e^+e^-$ center-of-mass (CM)
frame. In order to improve the $\Delta E$ resolution, the invariant mass of the $\eta$ ($\etapr$) 
candidate is constrained to its world-average value~\cite{PDG}. 
Signal events typically peak at the nominal $B$-meson mass for $M_{\rm bc}$ and at zero for $\Delta E$. 
We retain events with $M_{\rm bc}>5.22\gevcc$ and $-0.20\gev<\Delta E<0.15\gev$ for further analysis.

The average number of reconstructed $B$ candidates per event is $1.1$. In events with multiple $B$ candidates,
we select the one having the smallest value of $\chi^2=\chi^{2}_{\etapr}+\chi^{2}_{K^{*}(892)^{0}}$, where
$\chi^{2}_{\etapr}$ and $\chi^{2}_{K^{*}(892)^{0}}$ are the vertex-fit quality measures for $\etapr$ and
$K^{*}(892)^{0}$ candidates, respectively. 
The probability to select the correct signal candidate is about 94\% after all selection criteria.

The dominant background arises from the $e^+e^-\to q\qbar$ continuum process, where $q$ denotes $u$, $d$, $s$ or $c$. 
To suppress this background, we employ a neural network~\cite{Neurobayes} combining the following six variables. 
We use the cosine of the angle in the CM frame between the thrust axis of the $B$ decay and all other reconstructed particles 
and a Fisher discriminant formed out of $16$ modified Fox-Wolfram moments~\cite{KSFW}. 
These two quantities distinguish the spherical topology of $B$ decay events from the jet-like continuum events.
As the $B$ meson has a finite lifetime, the separation along the $z$ axis between the signal $B$ vertex and that of 
the recoiling $B$ is used to separate signal from continuum events in which most of the particles 
originate from the IP. The expected $B$-flavor dilution factor that ranges from zero 
for no flavor tagging to unity for unambiguous flavor assignment, calculated using recoiling $B$ decay 
information~\cite{flavor_tag}, also helps in distinguishing signal from continuum background. 
Owing to the difference in spin configurations of the decay, some discrimination power is inherent in the 
distribution of the following two observables: 
the cosine of the angle between the $B$ flight direction and the $z$ axis in the CM frame, and the cosine of 
the angle between the daughter $\gamma$ and parent $B$ momenta in the $\eta$ rest frame.

The training and optimization of the neural network
are accomplished with signal and continuum Monte Carlo (MC) events. The signal sample is generated using the 
\textsc{EvtGen} program \cite{EVTGEN} based on a model of the two-body decay of a pseudoscalar to a vector and a pseudoscalar, 
that incorporates the effect of final state radiation.
The neural network output ($\nb$) lies in the range $[-1.0,+1.0]$, with the events near
$-1.0$ ($+1.0$) being more continuum (signal)-like. We apply a criterion $\nb>-0.3$ to substantially
remove continuum events. With this requirement, we retain about $91\%$ of signal while rejecting $82\%$
of the $q\qbar$ background. The remainder of the $\nb$ distribution has a strong peak near $+1.0$ for signal 
and hence is difficult to model with a simple function. Instead, we use the transformed quantity
\begin{eqnarray}
 \nbprim=\ln\left(\frac{\nb-\nblow}{\nbhig-\nb}\right),
 \label{eq_nbtrans}
\end{eqnarray}
where $\nblow=-0.3$ and $\nbhig=+1.0$, to improve the robustness of the analytical modeling.
 
To study potential backgrounds from $B$ decays, we use a mixture of generic and rare $B\Bbar$
MC samples. The former is dominated by decays induced by $b \to c$ transition with relatively large branching fractions 
while the latter consists of rare decays induced by $b \to u,d,s$ transitions.
The number of background events expected from both samples is quite small.
Some rare $B\Bbar$ backgrounds exhibit a peaking structure in the $M_{\rm bc}$ and $\Delta E$
distributions. 
The $B^+\to\etapr K^+$, $B^0\to\etapr K^0_S$ and $B^0\to\etapr K^+\pi^-$ decays might  
mimic our signal. The $\Delta E$ peak is expected to be shifted from zero in the first two 
decays because of the loss of final-state particles or particle misidentification.
To suppress their contributions, we reconstruct the 
$B^+\to\etapr K^+$ and $B^0\to\etapr K^0_S$ with each of these hypotheses 
and reject the event if the reconstructed $B$ meson has $M_{\rm bc}>5.27\gevcc$ and $|\Delta E|<0.20\gev$. 
From the study with a large-statistics MC sample, we expect about ten $B^+\to\etapr K^+$ and four $B^0\to\etapr K^0_S$ 
events before this rejection and only five and one, respectively, with it while keeping $99\%$ of signal events. 

Contributions from the $B^0 \to\etapr K^+\pi^-$ (nonresonant) decay cannot be suppressed 
with the above method as the final state is identical to signal. 
In the fit procedure (described later) to extract signal, we fix the nonresonant background yield to two events, which corresponds 
to a branching fraction of $3.0 \times 10^{-6}$, estimated using the MC sample.
For the validation of this expected number, we have checked the background contribution using experimental data in 
the mass sideband of $1.0\gevcc < M_{K^{*}(892)^{0}} <1.2\gevcc$, and later extrapolated into the region used 
for our analysis. 
The $M_{K^{*}(892)^{0}}$ distribution in the nonresonant background decay is obtained by assuming a phase-space model. 
The nonresonant background contribution in the full data sample is estimated to be $3 \pm 4$ events, which is equivalent to a branching 
fraction of $(4.7\pm5.4) \times 10^{-6}$ and consistent with the two events from the MC sample. 
The difference of expected nonresonant background yields between the two strategies is incorporated into the systematic uncertainty.

We perform an unbinned extended maximum likelihood fit to the $M_{\rm bc}$, $\Delta E$, $\nbprim$
and $\cos\theta_H$ distributions of candidate events to extract the signal yield. The helicity
angle $\theta_H$ is defined as the angle between the momenta of the daughter charged kaon and the 
parent $B$ meson in the $K^*(892)^0$ rest frame. 
From an ensemble test of many pseudoexperiments, we find that $\cos\theta_H$ plays an important role in 
disambiguating the signal and nonresonant components, especially when the expected signal yield is small.
We define a probability density function (PDF) for each event category $j$ (signal, continuum $q\qbar$, generic $B\Bbar$,
rare $B\Bbar$ and nonresonant background) as:
\begin{eqnarray}
 {\cal P}^i_j\equiv{\cal P}_j(M^i_{\rm bc}){\cal P}_j(\Delta E^{i}){\cal P}_j(\nbprimi){\cal P}_j(\cos\theta^i_H),
 \label{pdf_equation}
\end{eqnarray}
where $i$ denotes the event index. As the correlation between each pair of fit observables is 
found to be small, the product of four individual PDFs is used as a good approximation for the true PDF. The 
likelihood function used in the fit is
\begin{eqnarray}
 {\cal L} = \exp\left(-\sum_{j}N_j\right)\times\prod_{i}\left[\sum_{j}N_j{\cal P}^i_j\right],
 \label{maximum_likelihood_equation}
\end{eqnarray}
where $N_j$ is the yield for event category $j$. For the signal, the correctly reconstructed $B$ meson decays are
referred to as the right-combination (RC) component while the misreconstructed decays are denoted as the self-crossfeed
(SCF) component. They are treated distinctly in the fitter with a combined PDF $N_{sig}
\times[f\,{\cal P}_{\rm RC}+(1-f)\,{\cal P}_{\rm SCF}]$, where $N_{sig}$ is the total signal yield and
$f$ is the RC fraction fixed to the value (94.5\%) determined from MC simulations.

Table~\ref{PDF_table} lists the PDF shapes used to model the $M_{\rm bc}$, $\Delta E$, $\nbprim$ and
$\cos\theta_H$ distributions for each event category. The PDF distributions that are difficult to parametrize
analytically are modeled using MC events either as histograms or smoothed shapes obtained with a kernel density estimation algorithm 
(Keys)~\cite{Keys}. 

The yields for all event categories
except for the rare $B\Bbar$ and nonresonant components are allowed to vary in the fit. The relative
contributions of the rare $B\Bbar$ and nonresonant background categories are very small and thus fixed to their MC values
($1.2\%$ and $0.7\%$, respectively). 
All signal shape parameters are fixed during the signal extraction after correcting them for possible 
differences between data and MC simulations using a high-statistics control sample whose final states are 
similar to the signal.
For $M_{\rm bc}$ and $\nbprim$, $B^0\to\etapr\KS$ is used as the control sample. 
The $B^0\to\Dzb\rho^0$ decay with $\Dzb\to K^+\pi^-\pi^0$ and $\rho^0\to\pi^+\pi^-$ is used  
to estimate the $\Delta E$ correction factors as the ones obtained from $B^0\to\etapr\KS$ are not sufficiently accurate. 

\begin{table}[ht]
\begin{center}
 \caption{List of PDFs used to model $M_{\rm bc}$, $\Delta E$, $\nbprim$ and $\cos\theta_H$ for
 the event categories. G (2G), BifG (2BifG), CB, P$_i$, ARGUS, and Hist denote single (double) Gaussian, single (double) 
bifurcated Gaussian, Crystal Ball~\cite{CB}, $i$-th order Chebyshev polynomial, ARGUS function~\cite{Argus},
 and histogram, respectively.}
  \vspace{1.0mm}
  \begin{tabular*}{85mm}{@{\extracolsep{\fill}}lcccc}
   \hline
   \hline
   Component          & $M_{\rm bc}$ & $\Delta E$ & $\nbprim$ & $\cos\theta_H$\\
   \hline
   Signal (RC)        & CB           & CB+BifG       & 2BifG       & Hist\\
   Signal (SCF)       & Hist         & Hist        & Hist      & Hist\\
   Continuum $q\qbar$ & ARGUS        & P$_1$       & 2G        & Hist\\
   Generic $B\Bbar$   & ARGUS        & P$_2$       & BifG        & Hist\\
   Rare $B\Bbar$      & Hist         & Hist        & BifG        & Hist\\
   Nonresonant background        & Hist         & Hist        & Hist      & Hist\\
   \hline
   \hline
  \end{tabular*}
\label{PDF_table}
\end{center}
\end{table}

\begin{figure}[h]
 \begin{center}
 \includegraphics[scale=0.38]{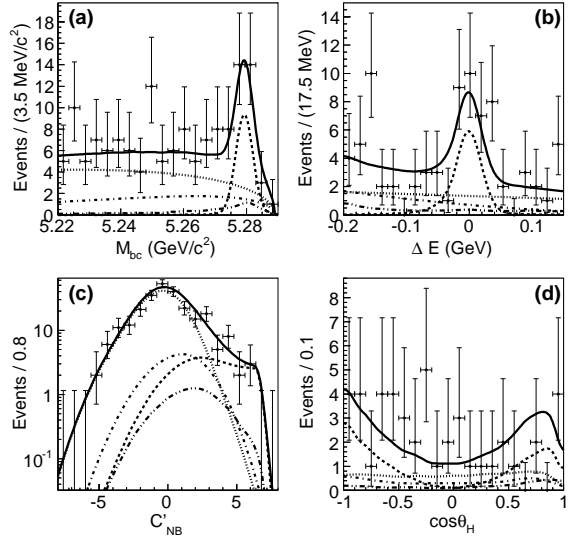}
 \caption{Projections of the fit results onto (a) $M_{\rm bc}$, (b) $\Delta E$,
 (c) $\nbprim$ and (d) $\cos\theta_H$. Each distribution is shown in the signal enhanced
 regions of the other three observables: $M_{\rm bc}>5.27\gevcc$, $-0.10\gev<\Delta E<0.06\gev$ and $2.0 < \nbprim < 8.0$. 
 Data are points with error bars; the fit results are
 shown by solid curves. Contributions from signal, continuum $q\qbar$, generic $B\Bbar$
 and rare $B\Bbar$ including nonresonant background are shown by dashed, dotted, dash-dotted, dash-double-dotted
 curves, respectively.
}
 \label{4dfit_to_real_data}
 \end{center}
\end{figure}

\begin{figure}[h]
\begin{center}
 \includegraphics[scale=0.45]{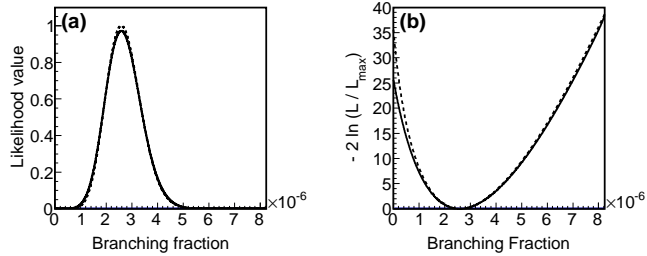}
 \caption{Distributions of (a) fit likelihood and (b) $-2\ln({\cal L}_0/{\cal L}_{\rm max})$ as
 a function of the branching fraction. Solid curves are after taking the systematic uncertainty into account 
 while dashed ones are only with the statistical uncertainty.}
 \label{NLL}
\end{center}
\end{figure}

Figure~\ref{4dfit_to_real_data} shows the $M_{\rm bc}$, $\Delta E$, $\nbprim$ and $\cos\theta_H$
projections of the result of the fit to data. We obtain $31\pm9$ signal, $2564\pm95$ continuum
$q\qbar$, and $253\pm82$ generic $B\Bbar$ events.  From the extracted yields, 
we obtain a significance of 6.0$\sigma$, where the significance
is defined as $\sqrt{-2\ln({\cal L}_0/{\cal L}_{\rm max})}$ with ${\cal L}_{\rm max}$ (${\cal L}_{0}$)
being the likelihood value when the signal yield is allowed to vary (fixed to zero). 
We calculate the branching fraction ${\cal B}[B^0\to\etapr K^{*}(892)^{0}]$ as 
\begin{eqnarray}
 {\cal B} &=& \frac{N_{sig}}{2 \times N_{B^0\Bbar^0}\times \varepsilon_{\rm rec}\times\varepsilon_{\rm PID}\times\varepsilon_{\nb}}\\
          &=& [2.6\pm0.7\stat\pm0.2\syst]\times10^{-6} \nonumber,
 \label{eq_BF}
\end{eqnarray}
where $2 \times N_{B^0\Bbar^0}$ is the total number of $B^0$ and $\Bbar^0$ ($772\times10^6$), $\varepsilon_{\rm rec}$
($1.7\%$) is the signal reconstruction efficiency including all daughter branching fractions,
$\varepsilon_{\rm PID}$ is a correction to the efficiency that takes into account the difference
between data and MC simulations for pion and kaon identification ($94.0\%$), and $\varepsilon_{\nb}$
is a similar correction factor for the continuum suppression requirement ($98.5\%$). 
Figure~\ref{NLL} shows the statistical significance convolved with a Gaussian function of width equal 
to the systematic uncertainty. In the significance calculation, we consider 
additive systematic uncertainties that affect only the extracted signal yield. There are also  
multiplicative uncertainties for all efficiency terms and the number of $B^0\Bbar^0$ pairs [Eq. \ref{eq_BF}]. 
The total significance after taking the systematics into account is $5.0\sigma$. 

In addition to the decay branching fraction, we also measure the $\CP$ violation asymmetry (${\cal A}_{\CP}$) 
by splitting the obtained yields according to the flavor of the decaying $B$ meson, based on the charge of 
the daughter kaon from the $K^*$ decay. 
From $N[\Bbar^{0}\to\etapr \overline{K}^*(892)^0]=12\pm6$ and $N[B^{0}\to\etapr K^*(892)^0]=19\pm6$, 
we obtain ${\cal A}_{\CP}$ for the decay as
\begin{eqnarray}
\hspace*{-6.0mm} {\cal A}_{\CP}&=&\frac{N[\Bbar^0\to\etapr \overline{K}^*(892)^0]-N[B^0\to\etapr K^*(892)^0]}{N[\Bbar^0
 \to\etapr \overline{K}^*(892)^0]+N[B^0\to\etapr K^*(892)^0]}\\
              &=&-0.22\pm0.29\stat\pm0.07\syst\nonumber,
 \label{eq_Acp}
\end{eqnarray}
where $N[B^0/\Bbar^0 \to\etapr K^*(892)^0/\overline{K}^*(892)^0]$ are the event yields obtained for the corresponding decays.

\vspace*{-1.0cm}

We enumerate the sources of systematic uncertainties for the branching fraction and ${\cal A}_{\CP}$
in Tables~\ref{sum_systematic} and \ref{sum_systematic_Acp}, respectively. 
The uncertainties due to PDF shape parameters are estimated by varying all fixed parameters within their uncertainties. 
To assign a systematic uncertainty for the fixed histogram PDFs, we perform a 
series of fits with the contents of each histogram bin fluctuated according to a Poisson distribution. 
The uncertainties due to the calibration factors used to correct for the signal PDFs are obtained by varying
the factors by their uncertainties. 
We calculate the uncertainty due to the fixed SCF fraction by varying the latter by $\pm50\%$. 
The uncertainties that arise from the fixed yield of rare $B\Bbar$ component are obtained by varying each of 
the fractions by $\pm50\%$. 
The fit bias is evaluated by performing an ensemble test comprising 300 pseudoexperiments, where the signal,
rare $B\Bbar$ and nonresonant background components are picked up randomly from the corresponding MC
samples and the PDF shapes are used to generate events for other categories. 
Due to limited MC statistics, we assign $0.8\%$ uncertainty on the absolute scale of the efficiency. 
The uncertainty due to the data-MC discrepancy for continuum suppression is obtained using the control 
sample of $B^0\to\etapr \KS$. We compare the results of two cases: one with the same $\nb$ requirement 
as for signal and the other without any requirement. The difference is then 
incorporated as a systematic error. 
The decay $B^0 \to \overline{D}^{0} \rho^0$, $\overline{D}^0 \to K^+\pi^-\pi^0$, in which final state particles are 
common to signal, is used to determine the systematic uncertainty associated with the $\varepsilon_{\rm PID}$ 
requirement and, for the $CP$ measurement, that due to detector bias. 
The systematic uncertainty of the $\eta$ reconstruction efficiency is calculated by
comparing data-MC differences of the yield ratio between $\eta\to 3\pi^0$ and $\eta\to\gamma\gamma$.
We use partially reconstructed $D^{*+}\to D^0(\KS\pi^+\pi^-)\pi^+$ decays to obtain the uncertainty
due to charged-track reconstruction ($0.35\%$ per track). Finally, we calculate the total systematic
uncertainty by adding all contributions in quadrature.
\begin{table}[h]
\begin{center}
  \caption{Summary of the considered systematic uncertainties for the branching fraction. The upper (lower) part of the table shows the additive and multiplicative uncertainties as described in the text.}
  \vspace{1.0mm}
  \begin{tabular*}{85mm}{@{\extracolsep{\fill}}lcc}
    \hline
    \hline
    \multicolumn{1}{c}{Source} &  \multicolumn{1}{c}{Uncertainties (\%)}\\
    \hline
    Signal PDF                         & $\pm$2.2\\
    $q\qbar$ PDF                       & +0.7 $-$0.9\\
    Generic $B\Bbar$ PDF               & $\pm$1.1\\
    Rare $B\Bbar$ PDF                  & +0.4 $-$0.5\\
    Histogram PDF                      & $\pm$0.7\\
    $M_{\rm bc}$ PDF shape calibration & +1.2 $-$1.5\\
    $\Delta E$ PDF shape calibration   & +1.1 $-$0.8\\
    $\nbprim$ PDF shape calibration    & +2.4 $-$2.6\\
    SCF fraction                       & +2.3 $-$2.2\\
    Rare $B\Bbar$ fraction             & +2.5 $-$2.6\\
    Nonresonant background fraction               & $\pm$2.9\\
    Fit bias                           & $\pm$2.8\\
    \hline
    MC statistics                      & $\pm$0.8\\
    $\varepsilon_{\nb}$                & $\pm$2.1\\
    $\varepsilon_{\rm PID}$            & $\pm$3.4\\
    $\eta$ reconstruction              & $\pm$1.5\\
    Track reconstruction               & $\pm$1.4\\
    $N_{B^0\Bbar^0}$                   & $\pm$1.4\\
    \hline
    Total                              &+8.1 $-$8.2\\
    \hline
    \hline
  \end{tabular*}
  \label{sum_systematic}
\end{center}
\end{table}

\begin{table}[h]
\begin{center}
\caption{Summary of the considered systematic uncertainties for ${\cal A}_{\CP}$.}
 \vspace{1.0mm}
 \begin{tabular*}{85mm}{@{\extracolsep{\fill}}lcc}
 \hline
 \hline
 \multicolumn{1}{c}{Source} &  \multicolumn{1}{c}{Uncertainties}\\
 \hline
 Signal PDF                         & $\pm$0.013\\
 $q\qbar$ PDF                       & +0.001 $-$0.002\\
 Generic $B\Bbar$ PDF               & +0.005 $-$0.007\\
 Rare $B\Bbar$ PDF                  & $< \pm$0.001\\
 Histogram PDF                      & $\pm$0.006\\
 $M_{\rm bc}$ PDF shape calibration  & $\pm$0.003\\
 $\Delta E$ PDF shape calibration   & +0.005 $-$0.004\\
 $\nbprim$ PDF shape calibration    & +0.013 $-$0.009\\
 SCF fraction                       & +0.001 $-$0.002\\
 Rare $B\Bbar$ fraction             & +0.005 $-$0.004\\
 Nonresonant background fraction               & +0.004 $-$0.003\\
 Fit bias                           & $\pm$0.011\\
 Detector bias                      & $\pm$0.062\\
 \hline
 Total                              & +0.067 $-$0.066\\
 \hline
 \hline
\end{tabular*}
\label{sum_systematic_Acp}
\end{center}
\end{table}

In summary, we have measured the branching fraction of $B^0\to\etapr K^{*}(892)^{0}$ 
using the full $\Y4S$ data sample collected with the Belle detector. We
employ a four-dimensional maximum likelihood fit for extracting the
signal yield. Our measurement ${\cal B}[B^0\to\etapr K^{*}(892)^{0}]
=[2.6\pm0.7\stat\pm0.2\syst]\times10^{-6}$ constitutes the first observation of
this decay channel with a significance of 5.0$\sigma$. We have also measured the $\CP$ asymmetry ${\cal A}_{\CP}[B^0
\to\etapr K^{*}(892)^{0}]=-0.22\pm0.29\stat\pm0.07\syst$, which is consistent with no $\CP$
violation.

We thank the KEKB group for the excellent operation of the
accelerator; the KEK cryogenics group for the efficient
operation of the solenoid; and the KEK computer group,
the National Institute of Informatics, and the 
PNNL/EMSL computing group for valuable computing
and SINET4 network support.  We acknowledge support from
the Ministry of Education, Culture, Sports, Science, and
Technology (MEXT) of Japan, the Japan Society for the 
Promotion of Science (JSPS), and the Tau-Lepton Physics 
Research Center of Nagoya University; 
the Australian Research Council and the Australian 
Department of Industry, Innovation, Science and Research;
Austrian Science Fund under Grant No. P 22742-N16;
the National Natural Science Foundation of China under Contracts 
No.~10575109, No.~10775142, No.~10825524, No.~10875115, No.~10935008 
and No.~11175187; 
the Ministry of Education, Youth and Sports of the Czech
Republic under Contract No.~LG14034;
the Carl Zeiss Foundation, the Deutsche Forschungsgemeinschaft
and the VolkswagenStiftung;
the Department of Science and Technology of India; 
the Istituto Nazionale di Fisica Nucleare of Italy; 
the WCU program of the Ministry of Education, Science and
Technology, National Research Foundation of Korea Grants
No.~2011-0029457, No.~2012-0008143, No.~2012R1A1A2008330,
No.~2013R1A1A3007772;
the BRL program under NRF Grant No.~KRF-2011-0020333,
No.~KRF-2011-0021196,
Center for Korean J-PARC Users, No.~NRF-2013K1A3A7A06056592; the BK21
Plus program and the GSDC of the Korea Institute of Science and
Technology Information;
the Polish Ministry of Science and Higher Education and 
the National Science Center;
the Ministry of Education and Science of the Russian
Federation and the Russian Federal Agency for Atomic Energy;
the Slovenian Research Agency;
the Basque Foundation for Science (IKERBASQUE) and the UPV/EHU under 
program UFI 11/55;
the Swiss National Science Foundation; the National Science Council
and the Ministry of Education of Taiwan; and the U.S.\
Department of Energy and the National Science Foundation.
This work is supported by a Grant-in-Aid from MEXT for 
Science Research in a Priority Area (``New Development of 
Flavor Physics'') and from JSPS for Creative Scientific 
Research (``Evolution of Tau-lepton Physics'').

\end{document}